\documentclass[aps,amsmath,amssymb,nofootinbib,preprintnumbers,showpacs]{revtex4}
\usepackage{epsfig}
\usepackage{mathrsfs,graphicx,bm,amsmath}

\newcommand{\STr}{\mathrm{STr}}

\newcommand{\epm}{\epsilon_{\text{M}}}

\def\di{\displaystyle}

\def\bg{\begin{eqnarray}\begin{array}{rcl}\displaystyle}
\def\eg{\end{array} &\di    &\di   \end{eqnarray}}
\def\bm#1{\begin{eqnarray}\begin{array}{#1}\di}
\def\bmo#1{\begin{eqnarray*}\begin{array}{#1}\di}
\def\bml#1#2{\begin{eqnarray}\begin{array}{#1}\label{#2}\di}
\def\bgo{\begin{eqnarray*}\begin{array}{rcl}\displaystyle}
\def\ego{\end{array} &\di    &\di \nonumber  \end{eqnarray*}}

\def\btensor#1#2{\renew\left#1\begin{array}{#2}\di}
\def\brtensor#1#2#3{\ren#3\left#1\begin{array}{#2}}
\def\botensor#1#2{\renew\left#1\begin{array}{#2}}
\def\etensor#1{\end{array}\right#1}

\def\STr{{\rm STr}}

\def\s0#1#2{\mbox{\small{$ \frac{#1}{#2} $}}}
\def\0#1#2{\frac{#1}{#2}}

\begin{document}

\preprint{HD-THEP-07-33}

\title{Three-body scattering from nonperturbative flow equations}

\author{S. Diehl${}^{b}$}
\author{H. C. Krahl${}^{a}$}
\author{M. Scherer${}^{a}$}

\affiliation{\mbox{\it ${}^a$Institut f{\"u}r Theoretische Physik,
Philosophenweg 16, D-69120 Heidelberg, Germany}\\
\mbox{\it ${}^b$Institute for Quantum Optics and Quantum
Information of the Austrian Academy of Sciences,}\\
\mbox{\it A-6020 Innsbruck, Austria}}

\begin{abstract}
We consider fermion-dimer scattering in the presence of a large positive scattering length in the frame of functional renormalization group equations. A flow equation for the momentum dependent fermion-dimer scattering amplitude is derived from first principles in a systematic vertex expansion of the exact flow equation for the effective action. The resummation obtained from the nonperturbative flow is shown to be equivalent to the one performed by the integral equation by Skorniakov and Ter-Martirosian (STM). The flow equation approach allows to integrate out fermions and bosons simultaneously, in line with the fact that the bosons are not fundamental but build up gradually as fluctuation induced bound states of fermions. In particular, the STM result for atom-dimer scattering is obtained by choosing the relative cutoff scales of fermions and bosons such that the fermion fluctuations are integrated out already at the initial stage of the RG evolution. 
\end{abstract}


\maketitle


\section{Introduction}

The scattering of three nonidentical fermions has been considered long time ago by Skorniakov and Ter-Matrirosian (STM) \cite{Skorniakov56}. For short-ranged interactions and a positive scattering length, the two-body sector supports a shallow bosonic bound state, the dimer. If the scattering length is large compared to the effective range, then the low energy physics becomes universal in the sense that all observables may be expressed in terms of a single length scale, the fermionic scattering length $a$. Making use of the insensitivity with respect to the short distance physics, STM extracted the universal ratio of fermion-dimer to fermion scattering length to be $a_3/a =1.18$. 

The three-body problem has a long history. In the 70s, Efimov extended the results for fermions to the bosonic case, where a short distance scale is needed to stabilize the particles, giving in turn rise to the Efimov effect, i.e. the existence of a sequence of three-body bound states \cite{Efimov70}. In subsequent papers, he revealed the universality of the three-body problem in systems with large scattering length \cite{Efimov70s}, and furthermore considered effective range corrections \cite{Efimov93}. 

The first derivation of the STM equation using Feynman diagrams was performed in \cite{Komarov64} using a purely fermionic formulation. In the 90s, the three-body problem was considered in the light of Effective Field Theory. The problem was reformulated in terms of an effective theory, where the dimer degree of freedom is implemented explicitly \cite{Bedaque97,Bedaque99}; for a review see \cite{Braaten04}.  Extensions of the standard STM problem in this framework include higher order corrections in the effective range \cite{Bedaque97} and higher partial waves \cite{Gabbiani00}. Recently, the problem has seen renewed interest, developing into an important nontrivial benchmark for new techniques. Those include nonperturbative methods like quantum mechanical approaches in position space \cite{Petrov103}, as well as perturbative techniques like the $\epsilon$ expansion around the critical dimension \cite{Rupak06}. 

In this paper, we address the problem of fermion-dimer scattering in the frame of nonperturbative functional renormalization group equations (FRG) for the effective action \cite{Wetterich:1992yh,Berges:2000ew,Aoki:2000wm}. This technique has been used successfully to quantitatively investigate critical phenomena, i.e. to analyze the universal physics at very long distances close to a phase transition \cite{Berges:1995mw}. It has also been applied for the study of complex many-body systems \cite{Baier:2003fw,Birse:2004ha,Diehl:2007th}, where typical length scales are set by the mean interparticle spacing and the inverse temperature. In this paper we demonstrate how it can be used to address nonperturbative fluctuation problems on even much shorter distances, i.e. the scattering of few particles. Our approach is based on a systematic vertex expansion of the effective action, which keeps the full momentum dependence of the one-particle irreducible (1PI) vertices.

Vertex expansion schemes have been worked out in various applications \cite{momentumdep}. In these works, the resulting flow equations exhibit a complexity which necessitates a full numerical treatment. Here we take advantage of the fact that for the few-body scattering problem at low energies, the complicated momentum and frequency dependence can be considerably simplified. This also allows for a direct comparison of our method to other schemes. Focusing on three-body (fermion-dimer) scattering here, we stress that the method is also suitable to address four-particle (dimer-dimer) scattering \cite{Kagan05,Gurarie06}, which will be discussed in a future publication. Furthermore, we point out that the diagrams resummed during the flow involve both inner fermion and boson lines. Situations like this also appear  in various many-body problems, e.g. in the analysis of quantum critical points where both fermions and bosons develop zero modes \cite{Hertz1976Millis1993}. The STM problem allows to study the simultaneous elimination of fermion and boson degrees of freedom in a relatively controlled setting. It may thus provide valuable hints for the future treatment of coupled fermion-boson theories in the frame of functional RG, as initiated in recent work \cite{Jakubczyk_Strack08}. 

This paper is organized as follows: In Sect. \ref{Method}, we briefly sketch the FRG method and formulate the problem in terms of a two-channel model for both a stable fermion field and a composite boson field, coupled via the Feshbach coupling $h_\varphi$. In the limit of large Feshbach couplings, universality emerges naturally as an infrared stable fixed point for the renormalized Feshbach coupling \cite{Sachdev06,Diehl:2007ri}, and our model becomes equivalent to a single channel model with point-like fermionic interactions. At this stage the problem is formulated for arbitrary density and temperature. In Sect. \ref{VacPro}, we then specify the prescription to project onto the physical vacuum of vanishing density and temperature, as appropriate for few-body scattering. The following section is devoted to a discussion of the (exact) solution of the two-body problem in the FRG framework. Sect. \ref{AtomDimer} then deals with fermion-dimer scattering: the FRG equation for the fermion-dimer vertex is derived and cast into a form closely resembling the STM integral equation. We then consider solutions of the RG equations for various choices of the relative cutoff scale at which fermions and dimers are integrated out relatively to each other, and show in which limit the STM result is reproduced. Conclusions are drawn in Sect. \ref{Conclusion}.

\section{Method and approximation scheme} 
\label{Method}

We study the scale dependence of the effective average action $\Gamma_k$ \cite{Wetterich:1992yh}, for reviews see \cite{Berges:2000ew,Aoki:2000wm}. It includes all fluctuations with momenta $q^2\gtrsim k^2$. In the limit $k\to 0$ where the averaging scale $k$ is removed, all fluctuations are included and $\Gamma_{k\to 0}$ approaches the full effective action. In
practice, the scale dependence is implemented by introducing suitable cutoff functions $R_k(q)$ in the inverse propagators. The dependence of $\Gamma_k$ on
$k$ obeys an exact flow equation,
\begin{eqnarray}\label{eq:FRG}
  \partial_k \Gamma_k &=& \frac{1}{2} \STr \,
  (\Gamma^{(2)}_k + R_k)^{-1}\,\partial_k R_k = \frac{1}{2} \STr\,\tilde \partial_k \,\log
  (\Gamma^{(2)}_k + R_k).
\end{eqnarray}
Here, the ``supertrace'' $\STr$ sums over spatial momenta $\vec q$ and Matsubara frequencies $\omega$ as well as over internal indices and species of fields, with a minus sign for fermions. The effective action is formulated in Euclidean spacetime. The second functional derivative $\Gamma_k^{(2)}$ represents the full inverse propagator in the presence of the scale $k$. Both $\Gamma_k$ and $\Gamma_k^{(2)}$ are functionals of the fields. For the last equation, the derivative $\tilde \partial_k$ is defined to act on the explicit scale dependence set by the cutoff function $R_k$, and not on the implicit scale dependence of $\Gamma_k$. This leads to a compact notation, and is advantageous to make direct contact with diagrammatic representations. In practice the above functional differential equation can only be solved approximately using a suitable truncation of the full effective action functional. The vertex expansion is obtained from expanding the last expression in powers of the fields.

The effective action is the generating functional of the 1PI $n$-point correlation functions. In the physical vacuum state of vanishing density and temperature, these objects can be directly related to the scattering amplitudes.

The effective action for fermions interacting via a Feshbach resonance can be described by a simple two-channel ansatz. In momentum space, and after analytical continuation to Euclidean frequencies ($\omega_{\text{M}}\to - i \omega$, where $\omega_{\text{M}}$ is the Minkowski frequency) it reads
\begin{eqnarray}\label{eq:Trunc}
  \Gamma_k &=&
  \int\limits_Q   \big[\psi^\dagger(Q)\big(\mathrm{i}\omega  +q^2 -\sigma_{\text{A}}\big)\psi (Q) + \varphi^*(Q) P_\varphi(Q)  \varphi (Q)\big]  \\
&& - \int\limits_{Q_1,Q_2,Q_3}h_\varphi\delta(Q_1-Q_2-Q_3)\big(\varphi^*(Q_1)\psi_1(Q_2) \psi_2(Q_3)- \varphi(Q_1)\psi^*_1(Q_2)\psi^*_2(Q_3) \big)\nonumber
\end{eqnarray}
with four-momentum $Q = (\omega, \vec q)$. 
Here $\psi = (\psi_1, \psi_2)$ represents the stable nonrelativistic fermionic atom field. We further introduce a composite boson field $\varphi$ which mediates the interactions between the fermions via the Yukawa or Feshbach coupling $h_\varphi$. The composite field can play various physical roles, depending on the region of parameter space under consideration \cite{Diehl:2007th}, as well as on the averaging scale $k$. Here we will restrict ourselves to the vacuum limit where the scales associated to many-body physics vanish ($T=n=0$), as well as to the limit of broad Feshbach resonances, where the value of the (bare) Feshbach coupling drops out as an independent scale in the problem, $h_\varphi \to \infty$. In this limit, the bosonic field is purely auxiliary at a high scale $k$ with no propagation, and acquires dynamics only in the limit $k\to 0$ by the virtue of fluctuations, as will be explained in more detail below. The only remaining scale in the problem is then the fermionic scattering length $a$. This realizes ``large'' scattering lengths, in the sense that the observable physics is uniquely determined by the latter value, becoming insensitive with respect to further microscopic information. This has an important aspect of universality, which is discussed for the few body-problem in \cite{Braaten04}, and with an emphasis on the implications for the many-body problem in \cite{Diehl:2005ae,Sachdev06,Diehl:2007ri}. The projection procedures on the physical vacuum and on the universal broad resonance limit are specified in the next section.

Our units are $\hbar = k_{\mathrm B} = c =1$. Furthermore, we measure all momenta in units of some reference scale $\hat k$, and energies in units of $\hat \epsilon_k = \hat k^2/2M$, where $M$ is the nonrelativistic mass of the fermions. This leads to a dimensionless scaling formulation of the effective action \cite{Diehl:2005ae}. We have the following relations between dimensionless and dimensionful (denoted with a hat) quantities \footnote{In an earlier publication \cite{Diehl:2005ae}, the dimensionless quantities were denoted with a tilde superscript, while the quantities without superscript were reserved for ``renormalized'' (rescaled with the wave function renormalization) quantities. We do not introduce such a wave function renormalization here, and it is advantageous to reserve the simplest notation for the dimensionless quantities.},
\begin{eqnarray}
Q = (\omega , q ) = (\hat \omega /\hat \epsilon_k , \hat q/\hat k), \quad \sigma_{\text{A}} = \hat \sigma_{\text{A}}/\hat \epsilon_k,  \quad h_\varphi = 2M \hat k^{-1/2} \hat h_\varphi ,\quad
P_\varphi (Q) = \hat P_\varphi (\hat Q)/\hat\epsilon_k, \quad \psi = \hat k ^{-3/2} \hat \psi, \quad \varphi = \hat k ^{-3/2} \hat \varphi .
\end{eqnarray}
The physical meaning of the parameter $\sigma_{\text{A}}$ again depends on the parameter regime under consideration. In the many-body context at finite density and temperature, it plays the role of the chemical potential for the fermions. In the vacuum, it represents \emph{half} the binding energy of a dimer (cf. Sect. \ref{UVRen}), which is nonzero for positive scattering lengths.

The inverse bosonic propagator consists of two contributions: a classical or ``bare'' part $\nu$ and a fluctuation part $\delta P_\varphi$,
\begin{eqnarray}
P_\varphi (Q) = \nu + \delta P_\varphi(Q).
\end{eqnarray}
The parameter $\nu$ defines the initial condition for the flow of the boson propagator. It includes the physical detuning $\nu (B)$ from the Feshbach resonance as well as a counter term $\delta\nu_{\text{in}}$ needed for the ultraviolet (UV) renormalization of the problem as discussed in Sect. \ref{UVRen}:
\begin{eqnarray}
\nu = \nu (B) + \delta\nu_{\text{in}}, \quad \nu = \hat \nu /\hat\epsilon_k , \quad  \nu (B) = \mu (B - B_0).
\end{eqnarray}
$\nu (B)$ measures the distance from the Feshbach resonance at magnetic field $B_0$, with $\mu$ being the effective magnetic moment of the atoms in the open channel. Dimensionless detuning and dimensionless scattering lengths (in the absence of an open channel background scattering length) are related by 
\begin{eqnarray}\label{Scattlength}
a  = - \frac{h_\varphi^2}{8\pi\nu (B)} , \quad a = \hat a \hat k.
\end{eqnarray}

The inverse fermion propagator 
\begin{eqnarray}
P_{\mathrm{F}} (Q) = \mathrm{i}  \omega  + q^2 - \sigma_{\text{A}}
\end{eqnarray}
does not receive any renormalization corrections in vacuum as will be shown in Sect. \ref{VacPro}. Therefore, we do not introduce scale dependent running couplings in the fermionic part of Eq. (\ref{eq:Trunc}). The running couplings for this part of the effective action are thus $h_\varphi$ and $P_\varphi(Q)$. 

In order to describe scattering processes involving more than two fermions, we need to extend the truncation Eq. (\ref{eq:Trunc}). In particular, the scattering amplitude of a fermion off a dimer is described by the amputated connected part of the Green function $\langle 0 | \varphi\psi \varphi^*\psi^\dagger |0 \rangle$ \cite{Braaten04}. Thus we need to include a fermion-dimer coupling 
\begin{eqnarray}
\int\limits_{Q_1, ... Q_4} \delta (Q_1+ Q_2-Q_3- Q_4) \delta \lambda_{3}(Q_1,Q_2,Q_3) \varphi(Q_1)\psi(Q_2) \varphi^*(Q_3)\psi^\dagger (Q_4).
\end{eqnarray}
The fermion fields are contracted as $\psi \psi^\dagger = \psi_\alpha \delta_{\alpha\beta} \psi^*_\beta$ in spin space, with $\delta_{\alpha\beta}$ the identity matrix in two dimensions. The order of $\psi$ and $\psi^\dagger$ is important due to the Grassmann nature of the fermionic fields and chosen such that it matches the standard conventions for the fermion-dimer scattering amplitude. The coupling depends on three independent four-momenta by momentum conservation. Still the four-momentum dependence is very involved and will be largely simplified below. However, a point-like truncation of the interaction vertices (no momentum dependence) turns out to be insufficient to reach satisfactory precision in this problem.  The dimensionless and dimensionful fermion-dimer couplings are related by 
\begin{eqnarray}
 \delta\lambda_{3} = 2M \hat k\,\delta\hat\lambda_{3}.
\end{eqnarray}

At this point we stress the systematic nature of the truncation advocated here. The vertex expansion is an expansion in the number of the fields. Our truncation (\ref{eq:Trunc}) is complete up to third order in the fields. At fourth order in the fields, there are two more terms which are compatible with $U(1)$ symmetry, namely
\begin{eqnarray} \label{OtherVert}
&&\int\limits_{Q_1, ... Q_4} \delta (Q_1- Q_2+Q_3- Q_4) \delta \lambda_{\psi}(Q_1,Q_2,Q_3) \psi^\dagger(Q_1)\psi(Q_2) \psi^\dagger(Q_3)\psi (Q_4),\\\nonumber
&& \int\limits_{Q_1, ... Q_4} \delta (Q_1- Q_2+Q_3- Q_4) \delta \lambda_{4}(Q_1,Q_2,Q_3) \varphi^*(Q_1)\varphi(Q_2) \varphi^*(Q_3)\varphi (Q_4).
\end{eqnarray}
The first one describes the scattering of two fermions, the second one the scattering of two dimers, i.e. four-fermion scattering in the presence of a bound state. The flow of these couplings has to be taken into account in a systematic expansion to fourth order in the fields. However, we will show in Sect. \ref{VacPro}:  (i) The first term is not generated by the flow in the vacuum limit considered here. This means that the fermionic two-body sector is fully described by our Feshbach model. (ii) The second vertex describes interactions in the four-body sector and is generated in the vacuum limit. In principle, it could couple into the flow of the other vertices. We find that this is not the case  in the vacuum limit. 

Finally, we specify the regulator functions $R_k$. We work with a momentum independent, mass-like cutoff function for fermions and bosons
\begin{eqnarray}\label{eq:cutoff}
R_{k,{\mathrm F}}=k^2\,,\quad R_{k,\varphi}= c k^2
\,,\end{eqnarray}
similar to \cite{Diehl:2007ri}. The choice of the dimensionless number $c$, which we specify below, sets the relative scale for the elimination of the fermionic and bosonic degrees of freedom in the renormalization group flow. The optimal choice of $c$ ensures an equal effective cutoff scale for fermions and bosons, where the effective cutoff scale is composed of the cutoff function piece plus possible physical mass terms \cite{Litim:2000ci, Litim:2001opt,Pawlowski:2005xe}.

A scheme mass-like cutoff function is possible if the fermionic ``chemical potential'' $\sigma_{\text{A}}<0$, which is indeed the case in the vacuum on the BEC side. For our purpose the mass-like cutoff is advantageous since it allows for most direct comparison with conventional diagrammatic techniques. However, for high accuracy calculation an optimized cutoff \cite{Litim:2000ci} would be more appropriate.


\section{Vacuum Limit}
\label{VacPro}

In this paper we consider a specific regime in parameter space where the effective action $\Gamma = \Gamma_{k=0}$ describes the scattering of particles in vacuum, which interact via a positive $s$-wave scattering length.  The vacuum projection of the effective action is obtained from $\Gamma$ in the limit $n\to0$, $T\to0$. 

The prescription, which projects the effective action on the vacuum limit reads \cite{Diehl:2005ae}:
 \begin{eqnarray}
 \Gamma_{vac} = \lim\limits_{k_{\mathrm F} \to 0} \Gamma_{k = 0} \Big|_{T > T_c }.
 \end{eqnarray}
Here $k_{\mathrm F} \equiv (3\pi^2 n)^{1/3}$ is directly related to the density of the system by definition, such that it can be viewed as the inverse mean interparticle spacing $k_{\mathrm F} \sim 1/d$. Taking the limit $k_{\mathrm F} \to 0$ then corresponds to a diluting procedure where the density of the system becomes arbitrarily low. However, the limit is constrained by keeping the dimensionless temperature $T = \hat T/\epsilon_{\mathrm F} = 2M \hat T/k_{\mathrm F}^2$ above criticality. This ensures that many-body effects such as condensation phenomena are absent. Of course, the system becomes arbitrarily cold, since the absolute temperature scales as $\hat T \propto k_{\mathrm F}^2 \to 0$.  
 
We find that for $n=T=0$ the crossover at finite density turns into a second-order phase transition in vacuum \cite{Diehl:2005ae,Sachdev06} as a function of  the magnetic field $B$. In order to see this, we consider the momentum independent parts in both the fermion and the boson propagator, $-\sigma_{\text{A}}$ (the ``chemical potential'' for the fermions in vacuum) and $m_\varphi^2$,  which act as gaps for  the propagation of fermions and bosons. Here the bosonic mass term is defined as the zero frequency and momentum component of the boson propagator,
\begin{eqnarray}
m_\varphi^2 = P_{k=0, \varphi} (Q = 0).
\end{eqnarray}
Taking the vacuum limit in the above mentioned form, we find the following constraints, which separate the two qualitatively different branches of the physical vacuum \cite{Diehl:2005ae},
\begin{eqnarray}\label{VacCond}
  \begin{array}{l l l}
    { m_\varphi^2 >0, \quad \sigma_{\text{A}} = 0  }& \text{atom phase}
      & (a^{-1} < 0) , \\
    { m_\varphi^2 = 0,\quad \sigma_{\text{A}} < 0   }& \text{molecule phase}
      & (a^{-1} > 0) ,  \\
    { m_\varphi^2 = 0,\quad \sigma_{\text{A}} = 0  }& \text{resonance}
      & (a^{-1} = 0).
  \end{array}
\end{eqnarray}
These formulae have a simple interpretation: On the BCS side, the bosons experience a gap $ m_\varphi^2>0$ and the low-density limit describes only fermionic atoms. On the BEC side, the situation is reversed: fermion propagation is suppressed by a gap $-\sigma_{\text{A}}$. The ground state is a stable molecule, and the fermionic chemical potential can be interpreted as half the binding energy of a molecule, $\epm = 2 \sigma_{\text{A}}$ \cite{Diehl:2005ae}, $- \sigma_{\text{A}}$ is the amount of energy that must be given to each fermion in a dimer to reach the fermionic scattering threshold.

Evaluating the flow equation for the bosonic mass term $m_\varphi^2$ with the constraint (\ref{VacCond}) on the BEC side (see below), one finds the well-known universal relation between binding energy and scattering length in vacuum, $\hat \epsilon_{\text{M}} = - 1/(Ma^2)$ in the broad resonance limit $\hat h_\varphi\to \infty$. This establishes the second order nature of the vacuum phase transition -- the resonance at $a^{-1}=0$ is smoothly approached. For finite $h_\varphi$, scaling violations $\mathcal O (\epm/h_\varphi^2)$ emerge \footnote{The situation is further complicated in the presence of an additional scale set by a finite background scattering length \cite{Diehl:2005ae}.}. This gives a glance at the status of universality related to the value of $h_\varphi$. 

On the technical side, the procedure specified above leads to a massive simplification of the diagrammatic structure as compared to the finite density and temperature system. By the aid of the residue theorem, it is straightforward to prove the following statement \cite{Diehl:2007ri}: All diagrams whose inner lines point in the same direction (thereby forming a closed tour) do not contribute to the flow in vacuum. Such diagrams have all poles in the same half of the complex plane. The argument holds for frequency and momentum dependent vertices provided that possible poles lie in the same half of the complex plane as those of the propagators. We can now analyze the one-loop diagrams which would possibly generate an RG flow of the couplings under consideration. An analysis of the one-loop diagrams is sufficient due to the one-loop structure of the exact flow Eq. (\ref{eq:FRG}). Applying the above statement, we find:\\
(i) The fermion propagator is not renormalized in vacuum as can be seen on diagrammatic grounds, and using the above argument. \\
(ii) A four-fermion vertex $\sim (\psi^\dagger\psi)^2$ is not generated by the flow. In our model, we have eliminated such a vertex to describe fermion-fermion scattering in favor of the coupling to the auxiliary boson degree of freedom (Hubbard-Stratonovich transformation). The fact that the vertex is not regenerated by the flow indicates that the Hubbard-Stratonovich transformation is very efficient here. We note that at finite density and temperature, such a vertex is indeed generated, describing the effect of particle-hole fluctuations.\\
(iii) The four-boson (dimer-dimer) vertex Eq. (\ref{OtherVert}) does not couple into the flow of the couplings we are considering in the frame of our truncation in vacuum. This implies that the four-particle sector of the theory does not affect the three-particle sector, which is physically sound. On the other hand, we find that the flow in the four-particle sector, described by the dimer-dimer vertex, is affected by the fermion-dimer vertex. The same pattern is observed for the mutual influence of the  two- and three particle sectors. Our vertex expansion therefore seems to respect the hierarchy which is expected from physical intuition.  At finite density, the interpretation of the vertices as representing scattering in sectors with definite particle number is spoiled. As expected, the many-body analogs of these vertices then do not respect the hierarchy any more. 


\section{Two-body sector: UV Renormalization and Universality}
\label{UVRen}

In order to make contact with experiment, we have to relate the microscopic or bare parameters which characterize the theory at a high momentum scale $k_{\text{in}}$ to the observables for two-atom scattering in vacuum, like the scattering length $a$, the molecular binding energy or an effective range. We therefore choose the initial parameters at the UV scale $k_{\text{in}}$ such that the two-body observables are matched in the limit $k\to 0$. 

In the two-body sector defined by Eq. (\ref{eq:Trunc}) we need to consider the flow of the Feshbach coupling $h_\varphi$ and the inverse boson propagator $P_{k, \varphi}(Q)$.
 
We consider the flow of the Yukawa coupling first. We find
\begin{eqnarray}
 \partial_k h_\varphi  = 0.
\end{eqnarray}
The non-renormalization of the Feshbach coupling in vacuum can be traced back to the $U(1)$ symmetry, and extends to the case of a momentum dependent Feshbach coupling. This statement holds in the absence of a fermionic background coupling. Extending the truncation to take such a coupling into account leads to a renormalization of $h_\varphi$ \cite{Diehl:2005ae,Diehl:2007ri}, which is compatible with charge conservation. 

\begin{figure}[t]
\centering
\includegraphics[width=60mm]{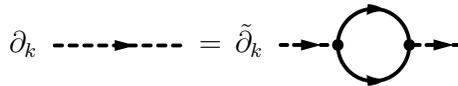}
\caption{\small{Graphical representation of the flow equation for the inverse bosonic propagator (cf. Eq. (\ref{BosPropI})) in the vacuum limit. Bosons (dimers) are denoted by dashed lines, fermions (atoms) by solid lines. Dots stand for the Yukawa coupling $h_{\varphi}$, which is not renormalized in vacuum.
}}
\label{fig:Bosepropkorr}
\end{figure}

The boson propagator is more involved and we discuss it in detail. The prescription projecting onto this object is given by the functional derivative
\begin{eqnarray}
\mathcal{P}_{k, \varphi} (Q_1, Q_2) = \frac{\delta^2}{\delta \varphi^* (Q_1) \delta \varphi (Q_2)} \bigg{|}_{\varphi=0,\psi=0}\,\, \Gamma_k .
\end{eqnarray}
We extract it by applying this prescription to both sides of Eq. (\ref{eq:FRG}), where in practice we expand the logarithm on the rhs in powers of the bosonic field. The inverse bosonic propagator is diagonal in momentum space, $ \mathcal{P}_{k, \varphi} (Q, K)=P_{k, \varphi} (Q) \delta(Q-K)$. This yields the flow equation (cf. Fig. \ref{fig:Bosepropkorr})
\begin{eqnarray}\label{BosPropI}
k\partial_k \delta P_{k, \varphi} (Q)  &=& - \int\limits_K k\tilde\partial_k\frac{h_\varphi^2}{\big(P_{\mathrm F}(- K)+k^2\big)\big(P_{\mathrm F}(K + Q)+k^2\big)}\nonumber\\
&=& \frac{h_\varphi^2 k^2}{8\pi}\frac{1}{ \sqrt{ \mathrm{i} \omega/2 + q^2/4 - \sigma_{\text{A}} +k^2  }}.
\end{eqnarray}
The flow is large ($\sim k$) for large cutoffs $k$. This reflects the presence of a relevant parameter and indicates the necessity of an UV renormalization in the language of the flow equation. To make the physics more transparent, we may consider the flows of the $Q\neq 0$ and $Q = 0$ (mass term) components separately,
\begin{eqnarray}\label{BosPropII}
k\partial_k [\delta P_{k, \varphi} (Q) -  \delta P_{k, \varphi} (0)] &=& \frac{h_\varphi^2 k^2}{8\pi}\Big(\frac{1}{ \sqrt{ \mathrm{i} \omega/2 + q^2/4 - \sigma_{\text{A}} +k^2   }} - \frac{1}{ \sqrt{ - \sigma_{\text{A}} +k^2  }}\Big) \sim \mathcal O (1/k),\nonumber \\
k\partial_k  \delta P_{k, \varphi} (0) &\equiv&  k\partial_k m_\varphi^2 = \frac{ h_\varphi^2}{8\pi}\frac{k^2}{\sqrt{-\sigma_{\text{A}}+k^2}}\sim \mathcal O (k).
\end{eqnarray}
Thus only the mass term $m_\varphi^2 = \delta P_{k, \varphi} (0)$ is UV sensitive, while the $Q \neq 0$ components are not. In the broad resonance limit $h_\varphi \to \infty$, we therefore find universality from the flow equations:  assuming initial conditions $\mathcal O (1)$, the loop contributions are $\mathcal O (h_\phi^2)$ and will therefore dominate the physical values of the couplings in the infrared limit, while memory of the initial conditions is lost. There is only a single relevant coupling, the mass term. This is the reason why we do not have to specify more details of the microscopic inverse boson propagator in Eq. (\ref{eq:Trunc}). For a more detailed discussion of universality in the frame of RG equations, we refer to \cite{Diehl:2007ri}.

As indicated in Sect. \ref{Method}, the initial condition for the inverse boson propagator is given by $P_{\text{in}, \varphi}(0) = \nu$ which defines the ``classical'' action for the boson degrees of freedom. Integrating the mass term in Eq. (\ref{BosPropII}) from the initial scale $k_\text{in} \to \infty$ down to the infrared limit $k =0$, we find the following relation
\begin{eqnarray}
\nu (B)  + \delta\nu_{\text{in}} - P_{0, \varphi} (0) =  \frac{h_\varphi^2}{8\pi} \,\, k_\text{in} - \frac{h_\varphi^2}{8\pi} \,\,\sqrt{- \sigma_{\text{A}}}.
\end{eqnarray}
The UV renormalization is thus performed by the choice $\delta \nu_{\text{in}} = h_\varphi^2/(8\pi) \,\, k_\text{in}$. Furthermore, we use the exact constraint Eq. (\ref{VacCond}) for positive scattering lengths, $P_{0, \varphi} (0) =0$, and conclude the relation (cf. Eq. (\ref{Scattlength}))
\begin{eqnarray}\label{ScattBind}
a = - \frac{h_\varphi^2}{8\pi \nu(B)}  = \frac{1}{\sqrt{-\sigma_{\text{A}}}} .
\end{eqnarray}
Since $-\sigma_{\text{A}}$ is the gap in the fermion propagator, it may be interpreted as \emph{half} the binding energy of a molecule, $\epm = 2\sigma_{\text{A}}: -\sigma_{\text{A}}$ is the amount of energy that has to be given to each of the fermions bound in a dimer in order to reach the scattering threshold. Therefore, we find the well known relation between scattering length and binding energy,
\begin{eqnarray}
\hat \epsilon_{\text{M}}  = 2\hat \sigma_{\text{A}} = -1/M\hat a^2. 
\end{eqnarray}
The scale dependent inverse boson propagator in vacuum is thus given by 
\begin{eqnarray}\label{ScBosProp}
\Gamma_{ k =0} ^{(2)}(Q) =  P_{k, \varphi} (Q) = \nu (B) +\delta P_{k, \varphi} (Q) = \frac{h_\varphi^2}{8\pi}\,\, \big(- a^{-1} +  \sqrt{ \mathrm{i} \omega/2 + q^2/4 - \sigma_{\text{A}} + k^2  }\big) .
\end{eqnarray}
In the presence of a nonzero binding energy ($\sigma_{\text{A}}<0$), we may expand the square root. Using Eq. (\ref{ScattBind}), we end up with
\begin{eqnarray}\label{LowBosProp}
\Gamma_{k=0} ^{(2)}(Q) \approx  \frac{h_\varphi^2}{32\pi\sqrt{-\sigma_{\text{A}}}} \,\,\big( \mathrm{i}\omega + q^2/2\big) = \epsilon_k^{-1} \hat h_\varphi^2 \frac{M^2 \hat a}{8\pi} \,\,\big( \mathrm{i}\hat \omega + \frac{\hat q^2}{4M}\big)
\,,\end{eqnarray}
which is the inverse Euclidean propagator for elementary bosons of nonrelativistic mass $2M$, dressed with a wave function renormalization 
\begin{eqnarray}\label{WFR}
Z_\varphi = \frac{h_\varphi^2}{32\pi\sqrt{-\sigma_{\text{A}}}} = \frac{h_\varphi^2 a}{32\pi}.
\end{eqnarray} 
The wave function renormalization coincides with the one obtained in \cite{Diehl:2005ae,Diehl:2007th,Diehl:2007ri} in the frame of a derivative expansion for the effective action.  Switching to Minkowski space we obtain the dimer dispersion $\hat \omega = \hat q^2/(4M)$. However, in the following calculations we use Eq. (\ref{ScBosProp}).

Let us summarize the results for the two-body problem. We have shown from diagrammatic arguments that the fermion propagator and the Feshbach coupling are not renormalized in vacuum. The renormalization of the boson propagator can be considered keeping the full frequency and momentum dependence. The flow is driven by fermionic diagrams only. We emphasize that the solution of the two-body problem is \emph{exact} as expected for point-like interactions. In our formalism, this is reflected by the fact that the two-body sector decouples from the flow equations for the higher order vertices: no higher order couplings enter Eq. (\ref{BosPropI}). Extending the truncation to even higher order vertices, or including the fermion-dimer vertex $ \varphi \psi \varphi^*\psi^\dagger$ does not change the situation, since there are only two external lines in the two-body problem and the flow equation involves only one-loop diagrams, such that contributions from such vertices cannot appear.

\section{Three-Body Sector: Atom-Dimer Scattering}
\label{AtomDimer}

\begin{figure}[t]
\centering
\includegraphics[width=70mm]{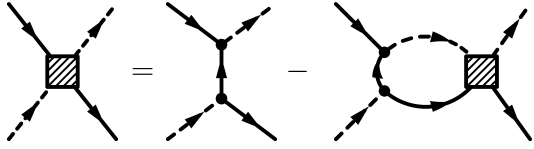}
\caption{\small{Graphical representation of the STM equation (cf. Eq. (\ref{STM})).
}}
\label{fig:STM}
\end{figure}

In this section we compute the fermion-dimer scattering amplitude $\lambda_{3}$ from which the fermion-dimer scattering length  $a_3$ can be extracted in the zero frequency and momentum limit. This problem has been formulated via a momentum space integral equation a long time ago by Skorniakov and Ter-Martirosian (STM) \cite{Skorniakov56}, for more recent treatments see \cite{Petrov103,Braaten04,Kagan05,Gurarie06}. The STM integral equation is derived from a consideration of possible scattering processes. These processes form a ladder structure and can thus be resummed in a Lippmann-Schwinger-type self-consistency equation, depicted in Fig. \ref{fig:STM}. 

Here, we present an alternative approach based on a first principles equation, the exact evolution equation for the effective action. We derive a flow equation for fermion-dimer scattering, and show how it relates to the STM result. Under certain assumptions, we can show the equivalence of both equations. The validity of these assumptions is checked via explicit numerical solution of the flow equation. 

The fermion-dimer vertex $\lambda_{3}$ is computed from
\begin{eqnarray}
\delta\lambda_{3} (Q_1, Q_2, Q_3)\delta (Q_1 + Q_2 - Q_3-Q_4) =\frac{\delta}{\delta \psi^*_1(Q_4)}\frac{\delta}{\delta \varphi^*(Q_3)} \frac{\delta}{\delta \psi_1(Q_2)} \frac{\delta}{\delta \varphi(Q_1)} \,\, \Gamma_k
\,.\end{eqnarray}
In the following we work in the center-of-mass (cm) frame. As our vacuum construction in Sect. \ref{VacPro} implies, we choose the boson to define the zero-point of energy, such that its four-momentum at rest in the cm frame is given by $P = (0,\vec 0)$. The (Minkowski) cm four-momentum of the fermion at rest reads $P = (-\sigma_{\text{A}}=-\epsilon_{\text{M}}/2,\vec 0)$, where $\epsilon_{\text{M}}$ is the dimensionless binding energy of the dimer -- the fermion propagator is gapped on the BEC side of the resonance. Our choice of the zero is different from the one of \cite{Kagan05,Gurarie06}, where the fermion energy is defined to be zero, such that the boson has negative energy $\epm = 2\sigma_A$. Of course such a shift in the zero of energy is arbitrary and our final equations are independent of this choice. With the cm four-momenta fixed, the effective dependence of the fermion-dimer amplitude is reduced to two independent four-momenta, and we will express this fact via the notation $\lambda_{3}(P_1,P_2; P)$. 

We consider the flow equation for the dimensionless fermion-dimer scattering vertex $\lambda_{3} $ for a specific set of external (Minkowski) four-momenta: Consider an incoming fermion with $P+P_1$, an incoming boson with $- P_1$. The outgoing momenta for the scattered fermion and boson can be written as $P+P_2$, $- P_2$. This configuration is in general off-shell \footnote{The on-shell condition reads $(\omega_{p_{1}} = \omega_{p_{2}}, |\vec p_1| = |\vec p_2|)$.}. As in the STM integral equation, the full off-shell amplitude is needed, since the fermion-dimer vertex also appears as a coupling in virtual processes described by one-loop expressions. 

\begin{figure}[t]
\centering
\includegraphics[width=140mm]{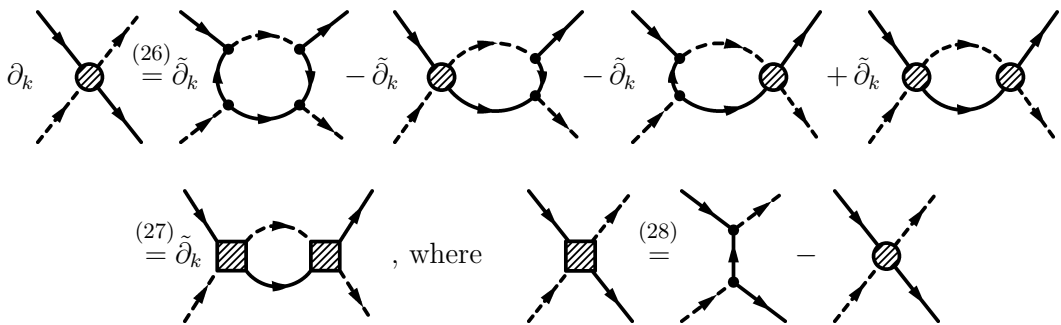}
\caption{\small{Graphical representation of Eq. (\ref{AtomDimerSimplified}). The shaded circles represent $\delta\lambda_{3}$, the shaded squares $\lambda_{3}$. The number of the corresponding equation in the text is displayed above the equality signs.
}}
\label{fig:floweq}
\end{figure}

We derive a flow equation for the frequency and momentum dependent fermion-dimer vertex as $\lambda_{3} (P_1,P_2; P)$. It is instructive to consider this equation in a form where the cutoff derivative on the rhs is not yet performed (arising from the expansion of the last expression in Eq. (\ref{eq:FRG})), since this allows for a direct comparison to standard diagrammatic techniques, see Fig. \ref{fig:floweq}. It reads \footnote{The flow equation is formulated in Euclidean space, while the physical frequencies are Minkowski frequencies. Therefore, we have to analytically continue these Minkowski frequencies $\omega_{\mathrm M}$ to Euclidean frequencies $\omega$ and insert these into the flow equation, where we use the relation $\omega = \mathrm{i} \omega_{\mathrm M}$.}
\begin{eqnarray}\label{eq:AtomDimer1}
\partial_k \delta\lambda_{3}(P_1,P_2; P) &=&   \int\limits_Q \tilde\partial_k\frac{1}{\big(P_{\mathrm F}(Q)+R_{k,\mathrm F}\big)  \big(P_{k, \varphi}(-Q+P)+R_{k,\varphi}\big)}  \Big[ \delta \lambda_{3}(P_1,Q;P) \delta \lambda_{3}(Q,P_2 ;P) \\\nonumber
&&\qquad \qquad -  \frac{h_\varphi^2}{P_{\mathrm F}(-Q - P_1)+R_{k,\mathrm F}} \delta \lambda_{3}(Q,P_2;P) 
 -  \delta \lambda_{3}(P_1,Q;P) \frac{h_\varphi^2}{P_{\mathrm F}(-Q - P_2)+R_{k,\mathrm F}}\\\nonumber
&&\qquad\qquad + \frac{h_\varphi^2}{P_{\mathrm F}(-Q - P_1)+R_{k,\mathrm F}}\,\frac{h_\varphi^2}{P_{\mathrm F}(-Q - P_2)+R_{k,\mathrm F}}\Big]\nonumber\\
&=&    \int\limits_Q \tilde\partial_k \frac{1}{\big(P_{\mathrm F}(Q)+R_{k,\mathrm F}\big)  \big(P_{k, \varphi}(-Q+P)+R_{k,\varphi}\big)}  \Big(\frac{h_\varphi^2}{P_{\mathrm F}(-Q - P_1)+R_{k,\mathrm F}} -\delta \lambda_{3}(P_1,Q;P)   \Big)\label{eq:AtomDimer2}\\\nonumber
&&\qquad\qquad \qquad \times   \Big( \frac{h_\varphi^2}{P_{\mathrm F}(-Q - P_2)+R_{k,\mathrm F}} - \delta \lambda_{3}(Q,P_2;P)  \Big).
\end{eqnarray}
We observe the emergence of the running coupling 
 \begin{eqnarray}\label{eq:Lambda3Full}
\lambda_{3}(K_1,K_2; P) =  \frac{h_\varphi^2}{P_{\mathrm F}(-K_1 - K_2) + R_{k,\mathrm F}} - \delta\lambda_{3}(K_1,K_2; P) 
\end{eqnarray}  
on the rhs of the above equation. The rhs thus exhibits a simple quadratic structure in the full vertex $\lambda_{3} = h_\varphi^2/P_{\mathrm F} - \delta\lambda_{3}$. However, the equation is not a closed equation for $\lambda_{3}$, since on the lhs only the induced coupling $\delta\lambda_{3}$ appears.

It is possible to make the comparison of Eq. (\ref{eq:AtomDimer1}) to the STM equation directly. However, for the sake of clarity and simple notation, we first simplify the above equation as appropriate for our purposes. This includes two main steps \cite{Kagan05,Gurarie06}: (i) Integrating out the loop frequency, and (ii) performing an $s$-wave projection, since we are only interested in low energy scattering here.

Step (i) is accomplished by noting that only the first fermion propagator $P_{\mathrm F}(Q)^{-1}$ appears with a positive sign for the loop momentum $Q$. For analytically continued (Euclidean) loop frequencies, there is consequently a single pole in the upper half plane, while the other propagators (as well as the vertex $\delta \lambda_{3}$, whose pole structure is generated by the tree contribution  $h_\varphi^2/P_{\mathrm F}$) are analytical in the upper half plane. We may thus perform the frequency integration by closing the contour in the upper half plane, implemented by setting the loop frequency $\omega \to - \mathrm{i} (q^2-\sigma_{\text{A}}+R_{k,\mathrm F})$ in the remaining momentum space integral. It follows that only values $\delta\lambda_{3}(P_1 = (p_1^2,\vec p_1), P_2 = (p_2^2, \vec p_2) ; P) \equiv \delta\lambda_{3}(\vec p_1, \vec p_2)$ are needed for the solution of Eq. (\ref{eq:AtomDimer1}). 

At this stage, Eq. (\ref{eq:AtomDimer1}) depends on six spatial momentum variables $\vec p_1, \vec p_2$. However, here we are only interested in low energy scattering. For microscopic short-range interactions described by the fermionic scattering length, scattering is dominated by the $s$-wave component \cite{Braaten04}. We may account for this fact by performing a suitable projection onto the $s$-wave. For this purpose, we average over the (independent) angles between both incoming and loop momentum, and outgoing and loop momentum by applying the integration over $\theta$: $1/2 \int_{-1}^{1}  d(\cos \theta)$. For the inner propagators, the averaging is performed explicitly. For the vertex, we define ($p_i = |\vec p_i|$)
\begin{eqnarray}
\delta\lambda_{3}( p_1,  p_2) = \frac{1}{2} \int\limits_{-1}^{1}  d(\cos \theta)\delta\lambda_{3}(\vec p_1, \vec p_2).
\end{eqnarray}

With these simplifications, Eq. (\ref{eq:AtomDimer2}) reduces to an equation which only depends on the two variables $p_1,p_2$. It reads
\begin{eqnarray}\label{AtomDimerSimplified}
\partial_k \delta\lambda_{3}(p_1,p_2)\!\! &=&\!\! \int  \frac{d q q^2}{2\pi^2} \tilde\partial_k\Big[\frac{8\pi}{h_\varphi^2\big( -a^{-1} +R_{k,\varphi}\frac{8\pi}{h^2_\varphi}+ \sqrt{\frac{3}{4} q^2 - \sigma_{\text{A}} +k^2+R_{k,\mathrm F}/2}\big)} \\\nonumber
&&\qquad\qquad \qquad \times    \Big(-\delta \lambda_{3}(p_1,q)  +\frac{h_\varphi^2}{4 p_1 q} \log\frac{ p_1 q + p_1^2 + q^2 - \sigma_{\text{A}} +R_{k,\mathrm F}}{- p_1 q + p_1^2 + q^2 - \sigma_{\text{A}} +R_{k,\mathrm F}} \Big)\\\nonumber
&&\qquad\qquad \qquad \times   \Big(-\delta \lambda_{3}(q,p_2)  +\frac{h_\varphi^2}{4 p_2 q} \log\frac{ p_2 q + p_2^2 + q^2 - \sigma_{\text{A}} +R_{k,\mathrm F}}{- p_2 q + p_2^2 + q^2 - \sigma_{\text{A}} +R_{k,\mathrm F}} \Big)\Big]\\\nonumber
&=&\!\! \int  \frac{d q q^2}{2\pi^2}\Big\{  \lambda_{3}(p_1,q) \Big[\tilde\partial_k\frac{8\pi}{h_\varphi^2( -a^{-1} +R_{k,\varphi}\frac{8\pi}{h^2_\varphi} + \sqrt{\frac{3}{4} q^2 - \sigma_{\text{A}} +k^2+R_{k,\mathrm F}/2})}\Big]  \lambda_{3}(q,p_2) \\\nonumber
&+&\!\!  \Big[ \tilde\partial_k \frac{h_\varphi^2}{4 p_1 q} \log\frac{ p_1 q + p_1^2 + q^2 - \sigma_{\text{A}} +R_{k,\mathrm F}}{- p_1 q + p_1^2 + q^2 - \sigma_{\text{A}} +R_{k,\mathrm F}} \Big] \frac{8\pi\lambda_{3}(q,p_2)}{h_\varphi^2( -a^{-1} +R_{k,\varphi}\frac{8\pi}{h^2_\varphi} + \sqrt{\frac{3}{4} q^2 - \sigma_{\text{A}} +k^2+R_{k,\mathrm F}/2})}\\\nonumber
&+&\!\! \frac{8\pi\lambda_{3}(p_1,q)}{h_\varphi^2( -a^{-1} +R_{k,\varphi}\frac{8\pi}{h^2_\varphi} + \sqrt{\frac{3}{4} q^2 - \sigma_{\text{A}} +k^2+R_{k,\mathrm F}/2})} \Big[\tilde\partial_k\frac{h_\varphi^2}{4 p_2 q} \log\frac{ p_2 q + p_2^2 + q^2 - \sigma_{\text{A}} +R_{k,\mathrm F}}{- p_2 q + p_2^2 + q^2 - \sigma_{\text{A}} +R_{k,\mathrm F}} \Big]
\Big\}
\,,\end{eqnarray}
where we have used Eq. (\ref{eq:Lambda3Full}) in the second step, and $-\sigma_{\text{A}} = a^{-2}$ (cf. Eq. (\ref{ScattBind})). Further we plug in the scale dependent solution for the inverse boson propagator obtained from integrating Eq. (\ref{BosPropI}). In the following it will be useful to represent Eq. (\ref{AtomDimerSimplified}) in matrix notation
\begin{eqnarray}\label{AtomDimerDiscret}
\partial_k \delta\lambda_{3} &=&   \lambda_{3}\cdot (\tilde\partial_k M) \cdot \lambda_{3} + 
 (\tilde\partial_k L) \cdot M \cdot \lambda_{3} + \lambda_{3}\cdot M  \cdot\tilde\partial_k L , \quad  \lambda_{3} = L - \delta\lambda_{3} , \end{eqnarray}
where the matrix elements are denoted by
\begin{eqnarray}
\lambda_{3}(q,q'),\!\!& & \\\nonumber
L(q,q') &=& \frac{h_\varphi^2}{4 q q'} \log\frac{ q q' + q^2 + q'^2 - \sigma_{\text{A}} +R_{k,\mathrm F}}{- q q' + q^2 + q'^2 - \sigma_{\text{A}}  +R_{k,\mathrm F}}, \\\nonumber
M(q,q') &=& \frac{8\pi dq q^2}{h_\varphi^2( -a^{-1} +R_{k,\varphi}\frac{8\pi}{h^2_\varphi} + \sqrt{\frac{3}{4} q^2 - \sigma_{\text{A}} +k^2 +R_{k,\mathrm F}/2})} \,\, \delta_{q,q'}
\, .\end{eqnarray}
The matrix $M$ is diagonal and includes the momentum space measure $d q q^2$. Furthermore, matrix multiplication means integration over a momentum variable \footnote{For the numerical solution we have discretized Eq. (\ref{AtomDimerDiscret}) on a logarithmically spaced two-dimensional momentum grid.}.

Eq. (\ref{AtomDimerSimplified}) represents a matrix equation which may be solved numerically. However, in the limit where the relative cutoff scale $c \to \infty$ (cf. Eq. (\ref{eq:cutoff})), the equation can be integrated analytically, with a solution that shows the equivalence to the STM equation in this limit. 

To see this most straightforwardly, we introduce a rescaled cutoff momentum $k' = c^{1/2} k$ and then draw the limit $c \to \infty, k' = \mathrm{const.}$ \footnote{In order to find $h_\varphi$-independent results appropriate for the broad resonance limit we need to scale $c\propto h_\varphi^2/(8\pi)$, similar to the detuning $\nu$.}
 The bosonic regulator then remains constant, while the implicit scale dependence of the boson propagator and the fermionic regulator are suppressed $\sim c^{-1}$. In particular, for $c \to \infty$, we have \begin{eqnarray}
\tilde \partial_{k'} M (c^{-1}) 
\to  \partial_{k'} M(0),\ L(c^{-1} ) \to L(0), \quad \tilde \partial_{k'} L(c^{-1} ) \to 0.
\end{eqnarray}
The scale derivative $\tilde \partial_{k'}$ reduces to a \emph{total} derivative in the limit $c\to \infty$. Eq. (\ref{AtomDimerSimplified}) thus reduces to $\partial_{k'} \delta \lambda_{3} = \lambda_{3} \cdot (\partial_{k'} M) \cdot \lambda_{3}$, which can be brought in a closed form for $\lambda_3$ since $L$ is $k'$-independent in the above limit, 
 \begin{eqnarray}\label{AtomDimerSol}
\partial_{k'} \lambda_{3} = -\lambda_{3} \cdot (\partial_{k'} M) \cdot \lambda_{3} , \quad \mathrm{or} \quad \lambda_{3}^{-1}\cdot (\partial_{k'} \lambda_{3}) \cdot  \lambda_{3}^{-1} = - \partial_{k'}  \lambda_{3}^{-1} =  - \partial_{k'} M.
 \end{eqnarray}  
This equation can be integrated straightforwardly with the result
\begin{eqnarray}\label{AtomDimerFinSol}
\lambda_{3} = (\textbf{1} + L\cdot M)^{-1}\cdot L. 
 \end{eqnarray}  
Here $\lambda_{3}$ denotes the full fermion-dimer vertex in the limit $k'\to 0$ where all fluctuations are included. Furthermore, the flow is initialized with $\lambda_{3,k' = k'_\text{in}} = L$, since for $k'=k'_\text{in}$ all virtual processes are suppressed and only the tree-level graph $L$ is present by construction of the flow equation.

At this point we observe that quite remarkably, the limit $c\to \infty$ leads to a solution of the flow equation, which is \emph{independent} of the choice of the cutoff function. All regularization scheme dependence drops out in this limit. This is because (i) the variables of the differential equation can be separated, and (ii) the function on the rhs of Eq. (\ref{AtomDimerSol}) becomes a total derivative. This property is lost if $c$ is taken finite. 

We are now in the position to directly compare this result to the STM integral equation \cite{Skorniakov56,Kagan05,Gurarie06}. Expressed in the above matrix notation it reads 
\begin{eqnarray}\label{STM}
\lambda_{3} = L - L \cdot M \cdot \lambda_{3}.
 \end{eqnarray}  
(For a graphical representation, cf. Fig. \ref{fig:STM}.) This equation is indeed solved by Eq. (\ref{AtomDimerFinSol}). 

Thus, we reproduce the STM equation in the limit where the relative cutoff scale $c \to \infty$. From the RG point of view, this limit implies that the fermions have already been integrated out at the UV scale where the flow is initialized. Both the bosonic propagator $P_{\varphi , k'} = P_{\varphi , c^{-1}k} \to P_{\varphi, 0}$ and the tree level graph $L_{k'} = L_{ c^{-1}k} \to L_{ 0}$ take their infrared values in the limit $c\to \infty$ already at the initial stage of the RG evolution. Both these quantities are obtained from the elimination of the fermionic degrees of freedom. Hence we recover the STM picture where both the boson propagator and the fermionic tree level graph are considered as  ``fundamental'' propagators and couplings, in the sense that they serve as the ingredients which allow to construct the Feynman diagrams of an effective theory describing fermion-dimer scattering. 

\begin{figure}[t]
\centering
\includegraphics[width=80mm]{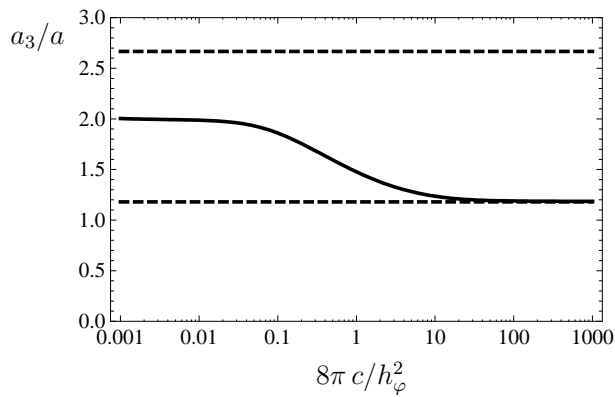}
\caption{\small{Dependence of the scattering lengths ratio $a_3/a$ on the relative cutoff scale $c$. For $c\to\infty$, the fermions are integrated out completely when the flow is initialized and the STM equation (lower dashed line, $a_3/a = 1.18$) is matched. Furthermore, the tree level scattering length ratio $a_3^{(\text{cl})}/a = 8/3$ is indicated (upper dashed line).
}}
\label{fig:CutoffDep}
\end{figure}

From the RG perspective it appears more natural that fermionic and bosonic degrees of freedom are integrated out simultaneously instead of first eliminating the fermions and then the bosons. This is especially true since the bosons are not fundamental but build up dynamically as fluctuation induced fermionic bound states. However, the fermions are gapped with half the binding energy $-\sigma_{\text{A}} = -\epsilon_{\text{M}}/2$, while the bosons become massless in the IR limit. In view of a judicious choice of $c$, we may use a simple optimization argument \cite{Pawlowski:2005xe}. The flow is initialized at $k^2 \gg -\sigma_A$. In this regime the cutoff function strongly suppresses the flow and the choice of $c$ is obviously arbitrary to a large extent. When $k^2 \gtrsim -\sigma_A$ we observe the onset of nonzero flow. Fermions and bosons couple to each other. The optimization sketched below Eq. (\ref{eq:cutoff}) requires equal effective mass terms for fermions and dimers $k^2 -\sigma_A \approx c k^2$. Thus $c \approx 1 -\sigma_A/k^2$ and we conclude a large optimal value of $c$ in the relevant flow regime $k^2 \lesssim -\sigma_A$. We emphasize that the large value of $c$ is due to the gap $-\sigma_A$ in the fermion propagator. This indicates that only small corrections can be expected once a full optimization program is implemented. This issue will be addressed in future work. Our simple prescription yields $a_3/a =1.22$. In Fig. \ref{fig:CutoffDep} we illustrate the impact of the choice of a $k$-independent relative cutoff scale on the result for the scattering length ratio $a_3/a$ for a wide range of $c$. A $c$-independent plateau is already reached for $8\pi c /h_\varphi^2 \approx 10$.  The shape of the curve in Fig. \ref{fig:CutoffDep} in the regime of small $c$ will depend on the choice of the cutoff functions, and only in the limit $c\to \infty$ do we obtain regularization scheme independent results. For finite $c$, further optimization -- by changing the \emph{shape} of the cutoff function \cite{Litim:2000ci, Litim:2001opt,Pawlowski:2005xe} -- will reduce the cutoff dependence, and presumably also lower the absolute value of $a_3/a$.

Finally, we specify the relation between the vertex function $\lambda_3(p_1,p_2)$ in the infrared limit $k\to 0$, and the fermion-dimer scattering length $a_3 \sim \lambda_3(0,0)$. The proportionality factor is obtained in the following way: First, the boson propagator has a nontrivial wave function renormalization $Z_\varphi = h_\varphi^2 a /(32 \pi )$ (cf. Eq. (\ref{WFR})). Absorbing it into the boson field to achieve a standard normalization of the frequency term, we find a renormalized vertex $\lambda_{3,R} = \lambda_{3}/Z_\varphi$. The dimensionful version of the vertex ($\hat \lambda_3 = \lambda_3/(2M \hat k)$) at zero momenta is related to the dimensionful fermion-dimer scattering length $\hat a_3$ by the standard relation, $\hat \lambda_3/Z_\varphi = (2\pi/M_\text{red})\,\, \hat a_3$, where $M_\text{red} = 2 M/3$ is the reduced mass of atom and dimer. Expressing these relations in dimensionless form, we get
\begin{eqnarray}
a_3 = \frac{ \lambda_{3,R}(0,0)}{ 6\pi }.
\end{eqnarray}


\section{Conclusions}
\label{Conclusion}

In this paper we have studied fermion-dimer scattering in the framework of FRG equations. In a systematic vertex expansion for the effective action, we derive the fully momentum dependent flow equations governing the two- and three-particle sector of the theory. The sectors form a hierarchy in the sense that the three-particle sector is determined by the two-body equations, but does not feed back into the latter. Our flow equations involve diagrams with inner fermion lines only as well as mixed diagrams, where both fermion and boson propagators appear in the loop diagrams.

We investigate in detail the relationship between the FRG equation for the fermion-dimer scattering vertex, and the STM integral equation.
In our flow equation approach, the starting point is a Yukawa-type theory for fermions and auxiliary, non-dynamical bosons after Hubbard-Stratonovic transformation on the level of the classical or microscopic theory. We then derive the flow equations for the two- and three particle sector of a theory for fundamental interacting fermions from first principles, i.e. a systematic vertex expansion of the exact flow equation for the effective action. In the two-particle sector, the dimer propagator is generated gradually in the RG flow. In the three particle sector, we encounter a flow equation which is diagrammatically equivalent to the STM equation. In our regularization scheme, we can choose the relative scale $c$ at which fermions and bosons are integrated out. Typically, fermionic and bosonic degrees of freedom are integrated out simultaneously in RG treatments, accounting for the fact that the dimers are not the elementary particles in the original purely fermionic theory.
In this work, however, we employ the limit  $c\to \infty$, where the fermions are integrated out completely already at the initial stage of the RG evolution. Then we can integrate the momentum dependent flow equation for the fermion-dimer scattering vertex analytically and show the equivalence to the STM equation, implying a ratio of fermion-dimer to fermion-fermion scattering length $a_3/a = 1.18$. We argue that the limit of large values for $c$ is physically sensible, since the fermions are gapped by half the binding energy, while the dimers become massless for positive scattering lengths (cf. Sect. \ref{VacPro}). We further show that the limit $c\to \infty$ is independent of the choice of the cutoff function. Thus the FRG treatment is suited to rederive the STM result correctly. We see this methodolocigal aspect as the main contribution of the present work.

Furthermore, the techniques developed in this paper can readily be generalized. One direction of future work is to investigate the effects of an effective range, obtained by considering smaller values for the Feshbach coupling $h_\varphi$. Furthermore, a fermionic background coupling, corresponding to scattering in the open channel in the atomic physics context, may easily be included in our framework.  A further goal is the treatment of the four-body problem in the presence of a positive scattering length, i.e., dimer-dimer scattering \cite{Petrov04,Kagan05,Gurarie06}. The 1PI four-body scattering amplitude $\langle \varphi\varphi^*\varphi\varphi^*\rangle$ is described by a dimer-dimer vertex and can be derived along the lines presented above. A preliminary analysis reveals that the fermion-dimer vertex feeds into the flow for dimer-dimer scattering. Taking this coupling into account produces the same diagrammatic topologies as found in the constructions \cite{Kagan05,Gurarie06}, which go beyond the resummation of the bosonic particle-particle ladder performed in \cite{AAAAStrinati,Diehl:2007ri}. Furthermore, the hierarchy of the sectors with different particle number is respected: The dimer-dimer coupling (4 particle sector) is affected by the couplings in the two- and three particle sectors, but does not feed back in the vacuum limit.

Our results are also relevant for a quantitatively accurate description of the BEC regime in the BCS-BEC crossover problem \cite{Sachdev06,Diehl:2005ae,Crossover}: As long as the binding energy of a dimer is the largest energy scale in the problem, the molecules should act as point-like entities. Therefore, the many-body physics is expected to be completely determined by the value of the dimer-dimer scattering length. As argued above, the dimer-dimer scattering length is in turn affected by fermion-dimer scattering.

Indeed, it has has been demonstrated recently that the value of the dimer-dimer scattering length in vacuum is directly reflected in many-body observables at low temperatures \cite{Diehl:2007th}. The formalism provided here is not bound to the physical vacuum, and effects of finite temperature and particle density can be straightforwardly included. Our technical developments are therefore useful for future studies of the BCS-BEC crossover, and may be seen as the first steps towards an approach which combines quantitative precision with a high degree of analytical insight into the crossover problem.

\emph{Acknowledgements} -- We acknowledge useful discussions with H. Gies, J. Pawlowski and C. Wetterich. SD also thanks E. Braaten, J. Braun, V. Gurarie, Yu. M. Kagan, I. Lesanovsky, J. Levinsen, A. Micheli, L. Radzihovsky and A. Schwenk for stimulating interactions. HCK acknowledges financial support by the DFG research unit FOR 723 under the contract WE 1056/9-1.



\end{document}